\documentstyle[amstex,amssymb]{article}

\begin{document}

\title{Spin-1 gravitational waves \\
and their natural sources}
\author{F. Canfora and G. Vilasi\\
\\
{\em Dipartimento di Fisica ''E.R.Caianiello'', Universit\`{a} di Salerno.}\\
{\em Istituto Nazionale di Fisica Nucleare, GC di Salerno, Italy. }}
\maketitle

\begin{abstract}
Non-vacuum exact gravitational waves invariant for a non Abelian
two-dimensional Lie algebra generated by two Killing fields whose commutator
is of light type, are described. The polarization of these waves, already
known from previous works, is related to their natural sources consisting of
cosmic strings and $\gamma$-ray bursts. Non vacuum exact gravitational waves
admitting only one Killing field of light type are also discussed.

\end{abstract}

\section*{Introduction}

Because of experimental efforts devoted to the detection of gravitational
waves, there is a growing theoretical interest in describing the emission of
detectable gravitational waves from astrophysical sources. All experimental
devices, laser interferometers or resonant antennas, need very precise
theoretical predictions, and usually they are calibrated according to
results of linearized theory. In fact, it is commonly believed that, since
the sources are very far from the earth, the wave's amplitude is so small at
the earth surface that the use of the linearized theory is lawful. However,
in some cases this assumption is misleading. Indeed, as shown by
Christodoulou \cite{Ch91} the nonlinearity of the Einstein equations
manifests itself in a permanent displacement of the test masses of a laser
interferometer which, in many physically interesting cases (e.g. for a
binary coalescence), cannot be neglected. Moreover, the nonlinearity of the
Einstein theory is one of its most characteristic feature, and it is likely
that some crucial properties of the gravitational field are encoded in the
nonlinear terms. Thus, a detailed analysis of exact gravitational waves,
besides to help the calibration of the experimental devices, could disclose
new phenomena hidden by the approximations.

In recent papers exact solutions \cite{SVV00} of vacuum Einstein field
equations, representing spin-1 gravitational waves \cite{CVV02,CVV02b}, have
been discussed. They are described by the following metrics:
\begin{equation}
g=2f\left( dx^{1}dx^{1}+dx^{2}dx^{2}\right) +\mu \left[ \left(
w(x^{1},x^{2})-2x^{4}\right) dx^{3}dx^{3}+2dx^{3}dx^{4}\right]  \label{svv1}
\end{equation}
where, $\mu =D\Phi +B$; $D,B\in {\cal R}$, $\Phi $ is a non constant
harmonic function of $x^{1}$and $x^{2}$, $\ f=\left( \nabla \Phi \right) ^{2}%
\sqrt{\left| \mu \right| }/\mu $, and $w\left( x^{1},x^{2}\right) $ is a non
constant\footnote{%
When $w$ is constant, metrics (\ref{svv1}) are static.} solution of the $\mu
$-{\it deformed Laplace equation:}
\begin{equation}
\Delta w+\left( \partial _{x^{1}}\ln \left| \mu \right| \right) \partial
_{x^{1}}w+\left( \partial _{x^{2}}\ln \left| \mu \right| \right) \partial
_{x^{2}}w=0,  \label{mdl}
\end{equation}
which, when $\mu $ \ is a constant function, reduces to the Laplace equation.

These metrics are intrinsically characterized \cite{SVV00} by the following
properties:

\begin{itemize}
\item  They are invariant for the non Abelian two-dimensional Lie algebra
generated by the Killing vector fields $X$ and $Y$ such that:
\[
\left[ X,\,\,\,Y\right] =Y,
\]

with $Y$ of {\it light-type}, i.e., $g\left( Y,Y\right) =0.$

\item  The distribution ${\cal D}^{\bot }$, orthogonal to the distribution$%
{\cal \ D}$ generated by $X$ and $Y$, $\ $is integrable and transversal to $%
{\cal D}$.
\end{itemize}

In the following, an integral (two-dimensional) submanifold ${\cal S}$ of $%
{\cal D}$ will be called a {\it Killing leaf} and an integral
(two-dimensional) submanifold ${\cal W}$ of ${\cal D}^{\bot }$ will be
called a {\it orthogonal leaf}.

In \cite{SVV00} it is shown that these Einstein manifolds have a have a
natural {\it fibered structure} with the Killing leaves as fibers and a
basis diffeomorphic to the orthogonal laves.

In the coordinates $\left( x^{\mu }\right) $, which are {\it adapted }\cite
{SVV00}{\it \ }to the Killing fields, the Killing vector fields $X$ and $Y$
read
\[
X=\frac{\partial }{\partial x^{3}},\,\,\,Y=\exp \left( x^{3}\right) \frac{%
\partial }{\partial x^{4}}.
\]

In the particular case $f=1/2$ and $\mu =1$, the above metrics are locally
diffeomorphic to a subclass of the vacuum Peres solutions \cite{Pe59}
corresponding to a special choice of the harmonic function parameterising
that metrics.

This is easily seen by introducing, for $u\neq 0$, new coordinates $\left(
x,y,u,v\right) $\ defined by
\begin{equation}
x^{1}=x,\,\ \ \,x^{2}=y,\,\ \ \ x^{3}=\ln \left| u\right| ,\,\ \ \ x^{4}=uv,
\nonumber
\end{equation}
in which metrics (\ref{svv1}) read

\begin{equation}
g=2f(x,y)\left[ dx^{2}+dy^{2}\right] +\mu (x,y)\left[ 2dudv+\frac{w}{u^{2}}%
du^{2}\right] .  \label{gp}
\end{equation}

A crucial problem in the theory of the gravitational waves is to
characterise realistic sources able to generate waves so strong to be
detected by the experimental devices. Thus, the aim of this paper is to find
physically interesting sources for the above metrics and to relate the
physical properties of such gravitational waves (in particular, the
asymptotic flatness, the wave-like character and the polarization) to the
sources. The plan of the paper is the following: in the first section, we
will solve the non-vacuum Einstein equations for metrics (\ref{svv1}) with a
dust source describing a null electromagnetic wave. In the second section we
will discuss the role of the asymptotic flatness and the related constraints
on the metric components and on the source. In this section we will also
discuss {\it cosmic strings} as a possible source. Then we will briefly
explain why the standard textbook analysis of gravitational waves does not
apply to this case. In the third section we will show that such
gravitational fields indeed represent gravitational waves by using two well
known covariant criteria: the Zel'manov's and the Pirani's criteria. In the
fourth section we will see that such waves are spin-1 objects. In the last
section we will give a hopefully vague discussion on the possibility to
detect such non standard gravitational waves.

\section{The dust and cosmic string sources{\it \ \ \ }}

The simplest source for metrics (\ref{svv1}) is {\it dust} with density $%
\rho $ and velocity{\it \ }$U^{\mu }${\it \ }and, then,{\it \ }characterized
by an energy-momentum tensor $T_{\mu \nu }=\rho U_{\mu }U_{\nu }$. When $%
U^{\mu }$ is a light-like vector field, this kind of energy-momentum tensor
can describe the energy and momentum of {\it null electromagnetic waves},
{\it i.e.,} electromagnetic fields whose scalars, $F^{\mu \nu }F_{\mu \nu }$
and $\epsilon ^{\alpha \beta \mu \nu }F_{\alpha \beta }F_{\mu \nu }$, both
vanish. In this way it could be possible to describe the gravitational
effects (in particular the emission of gravitational waves) of a very
interesting astrophysical phenomenon, the $\gamma -$ray bursts (GRBs):
emission of ultra-high energetic $\gamma $-rays, whose origin is still to be
fully understood.

Moreover, with this source one can also preserve the symmetries \cite
{SVV00,CVV02,CVV02b} of the vacuum solution.

Being the {\it time coordinate} in the Killing leaves, the dust cannot move
orthogonally to them and it will be chosen to move parallel to the {\it %
light-like} Killing field $Y$, {\it i.e., }with {\it velocity} $U^{\mu
}=\delta ^{\mu 4}$.

Thus, the metric will be taken in the form
\begin{eqnarray}
g &=&2F(x^{1},x^{2})\left( \left( dx^{1}\right) ^{2}+\left( dx^{2}\right)
^{2}\right)  \nonumber \\
&&+M(x^{1},x^{2})\left[ W(x^{1},x^{2},x^{4})\left( dx^{3}\right)
^{2}+2dx^{3}dx^{4}\right]  \label{m2}
\end{eqnarray}
with $F$, $M$ and $W$ \ to be determined by the non vacuum Einstein
equations with the following energy-momentum tensor
\begin{equation}
T_{\mu \nu }=M^{2}\rho \delta _{\mu 3}\delta _{\nu 3}.  \label{ms}
\end{equation}

Looking at the explicit expression of the Einstein tensor $G_{\mu \nu }$ it
turns out that
\[
G_{33}=WG_{34}+\frac{1}{2F}\left( M\Delta W+\nabla M\cdot \nabla W\right) ,
\]
where
\[
\nabla =\left( \frac{\partial }{\partial x^{1}},\frac{\partial }{\partial
x^{2}}\right) ,\quad \nabla M\cdot \nabla W=\frac{\partial M}{\partial x^{1}}%
\frac{\partial W}{\partial x^{1}}+\frac{\partial M}{\partial x^{2}}\frac{%
\partial W}{\partial x^{2}}.
\]
Then, since $T_{\mu \nu }=0$ for $\mu ,\nu \neq 3$, we can fulfil the non
vacuum field equations, while keeping the vacuum Killing fields, if and only
if $F=f$ and $M=\mu $, {\it i.e.} $F$ and $M$ coincide with their vacuum
expression \cite{SVV00}, and $W$ has the form $W\left(
x^{1},x^{2},x^{4}\right) \equiv w\left( x^{1},x^{2}\right) -2x^{4}$, $w$
being a solution of the $\mu $-{\it deformed Poisson equation}
\begin{equation}
{\Bbb \mu }\Delta w+\nabla {\Bbb \mu }\cdot \nabla w=2f{\Bbb \mu }^{2}\rho ,
\label{main}
\end{equation}
where, to save writing, beyond $c=1$, it has been taken $8\pi G=1$, $G$
being the Newton gravitational constant.

The comparison of Eqs. (\ref{mdl}) and (\ref{main}) shows that the
introduction of the matter source (\ref{ms}) induces only a simple and
natural modification in the field equations. The reason behind this
simplicity is that the above source, besides being physically interesting,
does not break the geometrical structures which allows \cite{SVV00} to
classify completely Ricci-flat metrics of the form (\ref{svv1}). In
particular, in this case it is still true that the $2$-dimensional
distributions orthogonal to the surface spanned by the Killing fields is
integrable since the energy-momentum tensor has no components orthogonal to
the Killing leaves. Thus, the bundle structure  \cite{SVV00} of the vacuum
space-time is not destroyed by the source.

\section{Asymptotic flatness \ \ }

From the physical point of view, it is important to characterise, among the
metrics (\ref{svv1}), those which are spatially asymptotically flat. The
vacuum case, where (for $\mu =1)$ $x^{1}$ and $x^{2}$ are harmonic
coordinates, suggests to call (spatially) {\it asymptotically flat} a metric
approaching, for $\left( x^{1}\right) ^{2}+\left( x^{2}\right)
^{2}\rightarrow \infty ,$ the Minkowski metric{\it .}

This intuitive definition of {\it asymptotic} {\it flatness} allows to
obtain qualitative results by using the standard theory of partial
differential equations.

In terms of the functions $f,$ $\mu $ and $w$, the asymptotic flatness
condition reads:
\begin{eqnarray*}
\left( x^{1}\right) ^{2}+\left( x^{2}\right) ^{2} &\rightarrow &\infty
\Rightarrow \\
f &\rightarrow &const,\quad {\Bbb \mu }\rightarrow const,\quad w\rightarrow
c_{1}x^{1}+c_{2}x^{2}+c_{3},
\end{eqnarray*}
where $c_{1},c_{2}$ and $c_{3}$ are arbitrary constants and the behaviour of
$w$ can be easily recognized by looking at the Riemann tensor.

It is easy to see that, in order for metrics (\ref{svv1}) to be spatially
asymptotically flat, necessarily $\mu $ must be constant. Indeed, being $f$
and ${\Bbb \mu }$ related by
\[
f=\frac{\sqrt{\left| \mu \right| }}{\mu }\left( \nabla \Phi \right) ^{2},%
\text{\thinspace \thinspace \thinspace }\mu =A\Phi +B,\quad \Delta \Phi =0,
\]
the condition $\mu \rightarrow const$ implies $A=0$ or $\Phi \rightarrow
const$. However, the case $\Phi \rightarrow const$ has to be excluded
because of the nondegeneracy of the metric.

For $\mu =1$, the equation for $w$ reduces to the two-dimensional Poisson
equation
\[
\Delta w=\rho .
\]
It is well known that, if $\rho $ goes to zero fast enough, it is possible
to find non trivial everywhere regular solutions $w$ tending to a constant
value\footnote{%
In the vacuum case this is not possible, unless $w=const$, because the
Laplace equation has not solution of this kind, as can be easily checked,
for example, by using the {\it maximum principle}.}.

The function $f$ satisfies the equation
\[
f\Delta f-\left( \nabla f\right) ^{2}=0,
\]
and this implies that the function
\[
\psi =\ln \left| f\right|
\]
is harmonic. Therefore, to have $f$ asymptotically constant, the function $%
\psi $ must be asymptotically constant. But this is impossible unless $\psi
=const$. Thus, in order to have everywhere regular spatially asymptotically
flat solutions, $f$ and $\mu $ must be constant functions and the fluid
density $\rho $ must tend to zero fast enough.

However, if we admit $\delta $-like singularities in the $\left(
x^{1},x^{2}\right) $ plane ({\it i.e.,} string-like singularities, by taking
into account the third spatial dimension), spatially asymptotically flat
vacuum solutions with $f\neq const$ and $w\neq const$ can exist. It is worth
noting that, in this limiting case in which $\rho \left( x^{1},x^{2}\right)
\rightarrow \delta \left( x^{1},x^{2}\right) $, the energy-momentum tensor
becomes the one usually employed to describe the gravitational effects of
topological defects known as cosmic strings \cite{VS00}. This kind of
extended objects are predicted in some {\it particles} {\it physics}
cosmological models with phase transitions. Moreover, {\it cosmic} {\it %
strings} could have an important role in the description of two very
interesting, but not yet fully understood, astrophysical phenomena: the GRBs
and the {\it ultra} {\it high} {\it energy} {\it cosmic} {\it rays}, i.e.
cosmic rays with energy $E\gtrsim 10^{11}GeV$, (for a review, see, for
example, \cite{Sa02}). It is important to stress here that {\it cosmic
strings} are not singular objects and should be described by a regular
energy-momentum tensor, but the above widely used scheme is useful for
practical computation.

Thus, from the phenomenological point of view, it is very intriguing that
these kind of (wave-like, as we will show later on) gravitational fields are
naturally coupled to cosmic strings. Eventually, unlike the electromagnetic
(or the standard gravitational) waves that can be everywhere regular even
without sources, these gravitational fields, to be globally nonsingular,
have to be coupled with matter sources. To be more precise, the reason why
in the standard textbook analysis of the polarization spin$-1$ gravitational
waves always appear as pure gauges is that one usually considers only {\it %
square-integrable} perturbations (in fact the Fourier analysis is needed to
kill such spin$-1$ components) to the Minkowski metric. However it is easy
to verify that, if one considers non {\it square-integrable} perturbations,
the spin$-1$ components cannot be gauged away anymore \cite{CVV02}. In the
case we are here considering, the physical components of the gravitational
field are not {\it square-integrable} and represent (as we will see later
on) spin$-1$ gravitational waves.

\section{Wave-character of the field \ \ }

From now on, to have a clearer physical interpretation, only the case $f=1/2$
and $\mu =1$ will be considered. In fact, in this case the metric (\ref{svv1}%
), besides being an exact solutions of the non vacuum Einstein equations, is
a solution of the linearized Einstein equations on flat background too.

As it has been said, metrics (\ref{svv1}) and (\ref{gp}) are locally
diffeomorphic. Moreover, for $f=1/2$ and $\mu =1$, the metric (\ref{gp}) is
a particular case of the Peres metric \cite{Pe59}:
\begin{equation}
ds^{2}=dx^{2}+dy^{2}+2dudv+h(x,y,u)du^{2}  \label{pere}
\end{equation}
where $h$ is an arbitrary function of $u$. If $h_{xx}+h_{yy}=0$, the metric (%
\ref{pere}) is a solution of vacuum Einstein field equation, otherwise,
since one still has $g^{\mu \nu }R_{\mu \nu }=0$ and $g^{\mu \nu }R_{\mu
\alpha }R_{\beta \nu }=0$, the source can be interpreted as a null
electromagnetic field \cite{MW57, Pe60}.

With the chosen energy-momentum tensor, the non-vacuum Einstein equations
for the metric (\ref{pere}) reduce to
\begin{equation}
\Delta h=\Delta \frac{w}{u^{2}}=\frac{\rho }{u^{2}}.  \label{peres}
\end{equation}

Now, let us briefly analyze the non vacuum Peres metric.

Firstly, it is trivial to show that $(x,y,u,v)$ are harmonic coordinates.
Furthermore, because of the Laplacian operator in Eq. (\ref{peres}), the
vacuum ({\it i.e.} with $\rho =0$) Peres solution also can be asymptotically
flat only if we admit $\delta -$like singularities in the $\left( x,y\right)
$ plane. In other words, the vacuum Peres solution is naturally coupled to
cosmic strings too. Eventually, it is easy to see that, in general, this
metric has only one\footnote{%
Of course, if $h=w/u^{2}$, then there exist an additional Killing field.}
(null) Killing field: $V=\partial /\partial v$ and that the non vanishing
independent components of the Riemann tensor are:
\[
R_{iuju}=-\partial _{ij}^{2}h,\quad i,j=x,y.
\]

The wave character and the polarization of these gravitational fields can be
analysed in many ways. For example, we could use the Zel'manov criterion
\cite{Za72} to show that these are gravitational waves and the
Landau-Lifshitz pseudo-tensor to find the propagation direction of the waves
\cite{CVV02, CVV02b}. However, the algebraic Pirani criterion is easier to
handle since it determines the wave character of the solutions and the
propagation direction both at once. Moreover, it has been shown that, in the
vacuum case, the two methods agree \cite{CVV02b}. To use this criterion the
Weyl scalars must be evaluated according to the Petrov-Penrose
classification \cite{Pet69, Pen60}.

To perform the Petrov-Penrose classification, one has to choose a {\it tetrad%
} basis with two real null vector fields and two real spacelike (or two
complex null) vector fields. Then, according to the Pirani's criterion, if
the metric belongs to type {\bf N} of the Petrov classification, it is a
gravitational wave propagating along one of the two real null vector fields.
However, the Peres coordinates are not of the above type and it is necessary
to introduce new coordinates adapted to the Petrov-Penrose classification.

By performing the following transformation
\[
x\mapsto x,\ \ \ \ y\mapsto y,\ \ \ \ u\mapsto u,\ \ \ \ \ v\mapsto
v+\varphi \left( x,y,u\right) ,
\]
whose Jacobian is equal to one,$\ $the metric (\ref{pere}) with the choice $%
h=\varphi _{,u}$ becomes:
\begin{equation}
g=dx^{2}+dy^{2}+2dudv+2(\varphi _{,x}dx+\varphi _{,y}dy)du  \label{spin1}
\end{equation}
and has, for generic $\varphi $, the only Killing field $Y=\partial
/\partial v$.

Since $\partial _{u}$ and $\partial _{v}$ are null real vector fields and $%
\partial _{x}$ and $\partial _{y}$ are spacelike real vector fields, the
above set of coordinates is the right one to apply for the Pirani's
criterion.

Since the only nonvanishing components of the Riemann tensor, corresponding
to the metric (\ref{spin1}), are
\[
R_{iuju}=-\partial _{ij}^{2}\partial _{u}\varphi ,\quad i,j=x,y,
\]
this gravitational fields belong to Petrov type {\bf N }(\cite{Ch84, Za72}).
Then, according to Pirani's criterion, the metric (\ref{spin1}) does indeed
represent a gravitational wave propagating along the null vector field $%
\partial _{u}$. It is worth noting here that, in order to have a spatially
asymptotically flat gravitational field, $\partial _{u}\varphi $ has to be a
solution of the two-dimensional Poisson equation in approaching a constant
at infinity; therefore, $\partial _{u}\varphi $, $\varphi _{,x}$ and $%
\varphi _{,y}$\ are not square-integrable functions.

\section{Polarization {\it \ \ }}

Once the propagation direction has been determined, to compute the
polarization we only need to look at the transformation properties of the
physical components of the metric (\ref{spin1}) under a rotation in the $%
\left( x,y\right) $ plane orthogonal to the propagation direction. By
''physical components'' we mean the true degrees of freedom of the metric.
Since the metric (\ref{spin1}) has only two independent components, we have
simply to look at the transformation properties\ of these components.
Moreover, Bel's superenergy tensor singles out the same degrees of freedom
\cite{CVV02b}. It is worth noting here that the metric (\ref{spin1}),
besides being an exact solution of the Einstein equations, is also a
solution of the linearized Einstein equations on flat background:
\begin{eqnarray*}
g &=&dx^{2}+dy^{2}+2dudv+2(\varphi _{,x}dx+\varphi _{,y}dy)du=\eta
_{ab}+h_{ab} \\
h_{ab} &=&\delta _{au}\delta _{bi}\varphi _{,i}
\end{eqnarray*}
where $\eta _{ab}$ is the flat metric and $h_{ab}$ can be considered as the
''perturbation''. Then, the spin may be discussed using the standard
textbooks analysis. However, before working out the transformation
properties of the physical components of the metric, we should verify that
the system of coordinate is harmonic (otherwise, there would be no
meaningful notion of polarization). In fact, by construction, the system of
coordinates of the metric (\ref{spin1}) is harmonic, so that the standard
procedure applies. Thus the analysis of the polarization of this
gravitational waves runs in exact the same way as in \cite{CVV02b}. The
result is the same: these are spin-$1$ gravitational waves because, roughly
speaking, the (non {\it square-integrable}) physical components of these
waves $h_{ux}=\varphi _{,x}$ and $h_{uy}=\varphi _{,y}$\ have only one index
in the $\left( x,y\right) $ plane orthogonal to the propagation direction $%
\partial _{u}$. It is worth noting that these results also hold when the
function $f$ in Eq. (\ref{svv1}) is not constant. The reason behind this
non-standard result could be that the spacetime topology is non trivial due
to the singularities of the function $\partial _{u}\varphi $ (as can be
easily recognized by considering the case in which $\partial _{u}\varphi $
is a solution of the Poisson equation with $\delta -$like sources, or, in
other words, when we consider the physical effects of cosmic strings). The
gauge transformations are able to kill spin-1 degrees of freedom only
locally, but in this case the singularities seem to be a global obstruction.

The possibility of spin$-1$ gravitational waves has also been guessed,
although in a rather different context, in \cite{ADK02} and \cite{KA02}.

\section{Possible experimental tests{\it \ \ \ }}

The following is a rather vague discussion on the possibility to detect such
non standard gravitational waves. The vagueness of the below comments is
mainly related to the lacking of experimental data on gravitational waves.
Nevertheless, we believe that it is interesting to outline in a very general
(vague) way how one could recognize the presence of spin-1 gravitational
waves in a would be gravitational wave's bursts detected by a laser
interferometer. The effects of these waves on neighbouring test particles
can be addressed by using the Jacobi {\it geodesic deviation} equation which
for the metric (\ref{spin1}) reads:
\begin{equation}
\frac{d^{2}}{d\tau ^{2}}Z^{i}=-g^{ih}Z^{j}\partial _{j}\partial _{h}\partial
_{u}\varphi ,  \label{js1}
\end{equation}
where $Z^{\mu }$ is a spacelike vector, with components only in the $\left(
x,y\right) $ plane, representing the displacement of two close\ geodetics
and $\tau $ is the proper time along the close geodesics.

The above equation shows that one can have either attraction or repulsion
according to the choice of $\varphi $, this $\varphi (x,y,u)$ being
constrained, outside the matter source, only by the condition to be a
harmonic function of $x$ and $y$. For example, the choice $\varphi =\rho
(x,y)\sigma (u)$, with $\rho $ a harmonic function\footnote{%
The choice $\rho =\ln \sqrt{x^{2}+y^{2}}$gives an asymptotic flat solution.}
of $x$ and $y$ and $\sigma $ a decreasing function of $u$, will give
repulsion. This is not surprising because it is known from QFT that spin-odd
bosons generate repulsion between particles of the same charge, the charge,
in this case, corresponding to the mass.

Now, the question is: ''Can we hope to detect such spin-1 waves?''. From the
theoretical point of view, it would be important to have an estimate of the
flux of spin-1 gravitational waves. However, it is rather difficult to
obtain even a rough estimate of the energy-density of all possible sources
of such waves. For this reason an approximate expression for $\rho $ is not
available yet, and, without $\rho $, it is not easy to estimate the expected
flux. From the experimental point of view, as it is well known, there are
delicate technical problems to detect gravitational waves from astrophysical
sources because of the weakness of the wave amplitudes with respect to
experimental noises. Nevertheless, once it will be possible to ''observe''
gravitational waves, it could be possible to distinguish the spin-2 and the
spin-1 components. Indeed, suppose to put the detector (such as a laser
interferometer or a resonant antenna) in the $\left( x,y\right) $ plane
orthogonal to the propagation direction of the waves. By definition, if a
spin-2 gravitational wave's burst is observed, then the experimental data
have to be invariant under a rotation of $\pi $\ in the $\left( x,y\right) $
plane. This means that the observable effects of the gravitational wave's
burst do not change if the detector is rotate of an angle equal to $\pi $.
However, if the burst contains also a spin-1 component, then the above
conclusion is not true anymore. In fact, the observable effects of the
spin-1 component are invariant under a rotation of $2\pi $ in the $\left(
x,y\right) $ plane. Therefore, a breaking of the invariance of the
experimental data under a rotation of $\pi $\ in the $\left( x,y\right) $
plane could be interpreted as the effect of spin-1 gravitational waves and
the amount of such a breaking could be used to get an estimate of the flux
of such waves. Moreover, since cosmic strings are naturally coupled to
spin-1 gravitational waves, if the device is able to detect mainly
cosmological gravitational waves, it could be possible, by looking at the
experimental data, to put further constraints on the presence of cosmic
strings. In fact, even a small spin-1 component could be the signature of a
non zero density of cosmic strings in the early cosmology.


\begin{thebibliography}{99}
\bibitem{ADK02}  D. V. Ahluwalia, N. Dadhich, M. Kirchbach, {\it Int. J.
Mod. Phys. }{\bf D} (2002)1621.

\bibitem{Ch84}  S. Chandrasekar, {\it The mathematical theory of black holes}%
, (Clarendon Press, Oxford, 1983).

\bibitem{Ch91}  D. Christodoulou, {\it Phys. Rev. Lett.} {\bf 67}, 1486
(1991).

\bibitem{CVV02}  F. Canfora, G. Vilasi, P. Vitale, {\it Phys. Lett. }{\bf B}%
, 545 (2002)373-378.

\bibitem{CVV02b}  F. Canfora, G. Vilasi, P. Vitale, {\it Spin-1
gravitational waves}, gr-qc/0212024; to appear in {\it Int.J.Mod. Phys. B}.

\bibitem{KA02}  M. Kirchbach, D. V. Ahluwalia, {\it Phys. Lett. }{\bf B},
529 (2002) 124-131.

\bibitem{MW57}  C. W. Misner, J. A. Weeler, {\it Ann. Phys. }(N.Y.) {\bf 2},
525 (1957)

\bibitem{Pe59}  A. Peres, {\it Phys. Rev. Lett.} 3, 571 (1959)

\bibitem{Pe60}  A. Peres, {\it Phys. Rev.} 118, 1105 (1960)

\bibitem{Pet69}  A. Z. Petrov, {\it Einstein spaces,} (Pergamon Press, New
York 1969)

\bibitem{Pen60}  R. Penrose, {\it Ann, of Phys.} 10:171, (1960)

\bibitem{Ro54}  N. Rosen, {\it Bull. Res. Coun. Isr.} {\bf 3}, 328 (1954).

\bibitem{Sa02}  M. Sakellariadou, {\it hep-ph}/0212365 (2002).

\bibitem{SVV00}  G. Sparano, G. Vilasi and A.M.Vinogradov, {\it Phys. Lett.}
{\bf B 513}, 142 (2001); {\it Diff. Geom. Appl.} {\bf 16}, 95 (2002); {\bf 17%
}, 1 (2002).

\bibitem{VS00}  A. Vilenkin, E. P. S. Shellard, {\it Cosmic Strings and
Other Topological Defects}, (Cambridge University Press, 2000).

\bibitem{Za72}  V. D. Zakharov, {\it Gravitational waves in Einstein's
theory,} (Halsted Press, N.Y.1973)
\end{thebibliography}
\end{document}